\documentstyle[12pt,aaspp4]{article}

\newcommand{\etal}{{\it et al.} }

\input psfig.tex
\begin{document}
\title{THE DISCOVERY OF 13\,s X--RAY PULSATIONS FROM THE HYDROGEN DEPLETED 
SUBDWARF O6 STAR BINARY HD49798}

\author{ G.L. Israel\altaffilmark{1,2}, L. 
Stella\altaffilmark{3,2}, L. 
Angelini\altaffilmark{4,5}, N. E. White\altaffilmark{4}, T. R.
Kallman\altaffilmark{4}, P. Giommi\altaffilmark{6} and A. 
Treves\altaffilmark{1}}

\altaffiltext{1}{International School for Advanced Studies (SISSA--ISAS), V. 
Beirut 
2--4, I--34014, Trieste, Italy, {Electronic mail: israel@sissa.it, 
treves@astmiu.uni.mi.astro.it}}

\altaffiltext{2}{Affiliated to the International Center for Relativistic 
Astrophysics}

\altaffiltext{3}{Osservatorio Astronomico di Roma, V. dell'Osservatorio 2, 
       I--00040 Monteporzio Catone (Roma), 
       Italy, {Electronic mail: stella@coma.mporzio.astro.it}}
       
\altaffiltext{4}{Laboratory for High Energy Astrophysics, Code 662, NASA -- 
Goddard 
Space Flight Center, Greenbelt, MD 20771, USA, {Electronic mail: 
white@adhoc.gsfc.nasa.gov, angelini@lheavx.gsfc.nasa.gov, 
tim@xstar.gsfc.nasa.gov}}

\altaffiltext{5}{University Space Research Association}

\altaffiltext{6}{SAX Science Data Center, ASI, Viale Regina Margherita 202, 
I--00198 Roma, Italy, {Electronic mail: giommi@sax.sdc.asi.it}}

\thispagestyle{empty}
\begin{abstract}
We discovered strong $\sim 13$\,s X--ray pulsations in the Rosat PSPC  
light curve of 
HD49798, a 1.55~d single--component spectroscopic  binary
containing a  hydrogen depleted subdwarf O6 star. We find no evidence for 
period changes during the $\sim 4$~hr Rosat pointing.  The source X--ray
spectrum is extremely soft, with an unabsorbed 0.1--2~keV  luminosity of a
few $\times10^{32}$~erg~s$^{-1}$  (distance of 650~pc). A higher luminosity
might be hidden in the EUV. Our results imply that the unseen  companion is an
accreting degenerate star, a white dwarf or, more likely, a neutron 
star. In either case HD49798 corresponds to  a previously unobserved
evolutionary stage of a massive binary system,  after common envelope and
spiral--in. 

\end{abstract}

\section{Introduction}

Subdwarf O stars form a fairly inhomogeneous group, spreading over a 
large range of temperatures, surface gravity 
and chemical compositions. 
Different evolutionary and nuclear histories probably contribute to the 
group of sdO stars, one of the late stages of stellar 
evolution that leads to the formation of white dwarfs 
(Bauer \& Husfeld 1995). 

The case of HD49798 is especially intriguing because this 8th magnitude 
sdO6 star, is also a spectroscopic binary (with P$_{orb}$ = 1.55~d,
$a \sin i$ = 3.60 R$_{\odot}$, f(m) = 0.263 M$_{\odot}$; Thackeray 1970; 
Stickland \& Lloyd 1994). Despite extensive studies, the nature of the 
companion star remained unclear for decades. Being 
very hot  (T$\simeq 47000$~K) and luminous (Log $(L/L_{\odot}) \simeq 3.90$) 
the sdO6 star outshines its companion 
(at least in the optical--UV). 
The level of H depletion, together with the underabundance of C the strong 
overabundance of N, testify that the outer layers were
processed  through the CNO cycle and the envelope of the progenitor 
star was lost during a common envelope phase. Kudritzki \& Simon (1978)
estimated a mass of $\sim 0.7-2.7$~M$_{\odot}$ for the sdO6 star and
suggested a similar  mass for the companion; they concluded that
HD49798 resulted from non--conservative mass transfer. 
Bisscheroux et al. (1996) argued that the sdO star has a
degenerate CO core and is in the phase of shell helium burning. 
Attempts at detecting the optical continuum 
of the companion provided conflicting conclusions:
Thackeray (1970) and Kudritzki \& Simon (1978) suggested 
a  F4--K0 main sequence companion;  
Goy (1978) favored instead a compact star. 
We report here the discovery of 13\,s pulsations in the 
soft X--ray flux from HD49798. 

\section{X--Ray Observations and Data Analysis}

The field of HD49798 was observed on 1992 Nov 11 from 18:07 to 22:06 UT 
(exposure of 5453\,s) with the Position Sensitive Proportional Counter 
(PSPC, 0.1--2.0~keV) in 
the focal plane of the X--ray telescope on 
board Rosat. 
Two X--ray sources were detected within $2^{\prime}$ 
from the center of the Rosat pointing. The Rosat error circle  
of the brightest of the two sources (1WGA~J0648.0--4418
center of RA = 06 48 04.6, DEC = --44 18 54.4  and
radius of $\sim 10^{\prime\prime}$, equinox 2000) includes the optical 
position of HD49798 ($\sim 5^{\prime\prime}$ offset).
The position of the $\sim~1.3^{\prime}$ distant, weaker X--ray 
source 1WGA~J0648.0--4419  (RA = 06 48 00.1, DEC =  --44 19 30.5) is 
inconsistent with HD49798.

The Rosat light curve and spectrum of HD49798 contained $\sim$~1000
photons  extracted from a circle of $1^{\prime}$ circle (containing to  
$>95$\% of the PSF) around the X--ray position. 
The contamination from the background and 
1WGA~J0648.0--4419 amounts to $\sim 20$ and $\sim 50$ photons, 
respectively. In order to correct for both we used the photons
extracted from a $1^{\prime}$ circle centered 
$\sim 1.3^{\prime}$ away from 1WGA J0648.0--4419 in the direction opposite to 
HD49798. By adopting different subtraction techniques we determined that 
only the PHA channels above $\sim 1$~keV are somewhat affected.  

The light curve of HD49798 was first analyzed as 
part of a study aimed at revealing  periodicities in  
$\sim 23000$ WGACAT X--ray sources (Israel \& Stella 1996; 
Israel \etal 1996). The photon arrival times were 
corrected to the barycenter of the solar system and a 
0.44~s binned light curve of HD49798 accumulated over the entire 
observation duration 
(3.7~hr). The corresponding power spectrum is shown in Fig.~1. 
The peaks around 0.0049 and 0.0125~Hz are seen in many 
Rosat sources and arise as a consequence of the wobble in the pointing 
direction (Israel \etal 1996). The 15$\sigma$ peak at 
$\sim 0.076$~Hz is instead unique to HD49798 
The best period was determined to be 
13.1789$\pm$0.0007~s by fitting the phases of the modulation 
obtained over 6 different intervals of $\sim 900$~s exposure 
(all uncertainties are 90\% confidence). 
The scatter of the phase residuals was consistent with a strictly periodic  
modulation at the best period 
($\chi^2$ of 4.7 for 4 degrees of freedom, $dof$). 
By adding a quadratic term to the fit a $3\sigma$ upper limit to the period
derivative of $|\dot{P}| < 2.3~\times~10^{-7}$~s~s$^{-1}$ was derived. 
The folded light curve is shown as an insert in Fig.~1. 
The arrival time of the pulse minima (adopted as phase 0)  
was JD~$2448938.332514\pm 0.000007$. 
The large pulsed fraction ($\sim 60\% $) 
and nearly sinusoidal shape of the modulation 
are consistent with being energy--independent in the PSPC band. 
The 13~s X-ray pulsations prove that HD49798 hosts 
a degenerate star accreting from the sdO star wind (see Sect.~3). 

The Rosat PSPC spectrum of HD49798 is extremely soft. 
Among single--component models, a power law produced by far the best fit 
($\chi^2/dof$ of 22.1/14)  
for a photon index of $\Gamma= 4.7\pm0.5$ and an interstellar column density 
of $N_H = (2.1\pm^{0.7}_{0.4})\times 10^{20}$~cm$^{-2}$ (see Fig.~2a). 
The corresponding 0.1--2.0~keV X--ray flux at the earth is 
$F \sim 8\times$10$^{-13}$~erg~cm$^{-2}$~s$^{-1}$. 
For the  estimated distance of 650~pc (Kudritzki \& Simon 1978) this converts 
to an unabsorbed luminosity of  $\sim 4\times$10$^{32}$~erg~s$^{-1}$. 
The extrapolation of this spectrum towards low energies provides a 
flux about two orders of magnitude lower than the IUE flux measurement
at 1200\AA\ ($\sim 2\times10^{-10}$~erg~cm$^{-2}$~s$^{-1}$~\AA$^{-1}$, 
Bohlin \etal 1990). This indicates that a luminosity of up to 
$\sim 10^{34}-10^{35}$~erg~s$^{-1}$ might be hidden in the EUV. 

Fig.~2a shows that a large contribution to the $\chi^2$ derives 
from a deficit of photons in the 0.4-0.6~keV range. 
Given the temperature, flux and peculiar composition of the 
sdO star atmosphere, HeII and the higher ionization stages of N 
(from NIV up) should dominate the wind optical depth for photon energies 
in the PSPC band
(see Sect.~3). This leads to a natural interpretation of the 
0.4-0.6~keV deficit in terms of K-edge absorption from N ions in the 
wind. By including an edge at 0.46~keV (NIV) the power law fit  provides a
$\chi^2/dof$ of $10.0/13$, for an optical depth of $\tau_N=3.4\pm^{3.0}_{1.9}$.
This gives an F--test probability of $1.5\times10^{-3}$ for the addition of 
one free  parameter. In
this case $\Gamma$ is $4.1 \pm 0.4$, 
while $N_H$ remains unchanged. Similar results are obtained 
if a NV K-edge at 0.50~keV is used. 
An upper limit to the HeII absorption in the wind  can be obtained
by ascribing the entire low energy absorption to these ions, {\it i.e.}
setting the interstellar absorption $N_H$ to zero; this gives
$\tau_{HeII} < 120$ for an edge at $54$~eV, with 90\% confidence. As 
discussed in Sect.~3, the N and He optical depths obtained above would be
reconciled if most of the He in the sdO star wind were fully ionized. 
Note that with the inclusion of an N-edge at 0.46~keV, a blackbody model 
provides also a reasonably good fit ($\chi^2/dof = 17.8/13$) for a 
temperature of $kT_{bb} = 114\pm12 $~eV and equivalent radius of 
$R_{bb}= 2.2\pm^{0.4}_{0.6} $~km; this gives a bolometric blackbody 
luminosity of $0.7-2 \times 10^{32}$~erg~s$^{-1}$. 

The Rosat PSPC spectrum of HD49798 can also be interpreted in terms of
the sum of a black body spectrum plus a high energy
excess, described by a power law 
(Israel et al. 1995). 
The lower $\chi^2/dof = 9.7/13$ resulting from the addition of 
two free parameters to the power law model gives an F--test 
probability of $7\times 10^{-3}$. 
The best fit is obtained for 
$kT_{bb}\simeq18.4$~eV, $\Gamma \simeq 3$ and 
$N_H \simeq 5\times 10^{20}$~cm$^{-2}$ (see Fig.~2c). 
The contribution of the high energy power law component is $\sim 30\%$.
In this interpretation the spectral paramaters are poorly constrained. 
This is clearly seen from Fig.\,2d where 
confidence contours are plotted in the $N_H$ -- $kT_{bb}$ plane. 
Only a lower limit of $N_H > 10^{20}$~cm$^{-2}$ 
and an upper limit of $kT_{bb} < 50$~eV can be deduced from the Rosat 
spectrum. 
A very wide range of black body luminosities is 
allowed, starting from $2\times 10^{32}$~erg~s$^{-1}$ (see Fig.\,2d).   
This is due to the fact that, 
for decreasing temperatures, a larger and larger fraction of 
the black body flux would be hidden in the EUV. 

The 1400~s {\it Einstein} HRI (0.05--4.0~keV) observation of HD49798 
of 1979 March 19 revealed an insufficient number 
of photons ($\sim 40$) to confirm the 13~s period. To within the 
uncertainties the Einstein HRI count rate is consistent with the Rosat PSPC 
measurements.

The EXOSAT CMA (0.05--2 keV) 
observed HD49798 on 1983 September 21 for a 
total of $\sim$7400~s. The source was observed at a count rate of 
$0.63\pm 0.04$, $0.40\pm 0.03$ and $0.035 \pm 0.005$~cts~s$^{-1}$, 
respectively with the polypropilene (PPL), 3000\AA\ and 
4000\AA\ Lexan filters. The ratio of these rates is incompatible with 
any plausible X-ray spectrum, clearly indicating a high level 
of UV contamination in the CMA data. The marginally significant  
$\sim 14.4$~s periodicity that we found in the PPL light curve is almost 
certainly introduced in the event processing by the on board computer
(similar to other CMA sources).

According to the ephemeris of Stickland \& Lloyd (1994), the Rosat 
and {\it Einstein} observations took place at orbital phase 
$0.00-0.10$ and $0.80-0.82$, respectively.
Therefore the presence of an X-ray eclipse (expected to be centered at 
$0.75$) cannot be ruled out yet.

\section{Discussion}

The discovery of 13~s pulsations in the X--ray flux of HD49798 proves that the
long--sought companion is a degenerate star, either
a white dwarf or a neutron star. The inferred radius of the sdO star is
$R_* \sim 10^{11}$~cm, about a half of its Roche Lobe.
Therefore, mass transfer towards the companion  must be driven
by the sdO star wind. Based on the 
the P--Cygni profile of the NV resonance doublet, Hamann \etal (1981)
determined that the terminal wind
velocity of $v_w\sim 1350$~km~s$^{-1}$ is reached at $\sim 1.7\ R_*$,
$i.e.$ within the Roche Lobe of the sdO star. The estimated mass loss rate
$\dot M_*$ ranges between
$5\times 10^{-10}$ and $10^{-8}$~M$_{\odot}$~yr$^{-1}$. 
Bisscheroux et al. (1996) argue that $\dot M_*$  
can be as high as  $\sim 3\times10^{-8}$~M$_{\odot}$~yr$^{-1}$
and the bulk of the wind matter (mainly H and He) 
as slow as $v_w\sim 800$~km~s$^{-1}$ 
By using the whole range of estimated $v_w$ and $\dot M_*$, and 
the standard theory of wind accretion in binary systems (see White, Nagase 
\& Parmar 1995 and references therein) 
we derive a mass capture rate of 
$\dot M_x \sim 10^{11}-6\times 10^{14}$~g~s$^{-1}$
by the degenerate companion. In deriving these values we used 
an orbital velocity of the degenerate star in the $80-160$~km~s$^{-1}$
range, as obtained from the measured mass function and velocity amplitude
($K\simeq 118$~km~s$^{-1}$) by allowing the degenerate and sdO star mass
to be $0.85-1.5$~M$_{\odot}$ and $0.7-2.1$~M$_{\odot}$, respectively.
(Note that the upper limit of $2.1$~M$_{\odot}$ 
is obtained from the mass function by using $i=90^{\rm o}$ and a 
maximum mass of $1.5$~M$_{\odot}$ for the degenerate star).
For the adopted mass range the system inclination is $i \geq 46^{\rm o}$. 

By using $\dot M_x$ given above, an accretion luminosity of
$2\times 10^{28}-2 \times 10^{32}$~erg~s$^{-1}$
is predicted in the case of a white dwarf and of
$1 \times 10^{31}-1\times 10^{35}$~erg~s$^{-1}$
in the case of a neutron star. While the former range is only marginally
consistent  with the inferred unabsorbed luminosity in the PSPC band 
(see Sect.~2), the second overlaps comfortably, allowing also 
for a larger luminosity in the EUV.  
By using the maximum angular momentum of the 
accretion flow relative to the companion, a circularization radius of 
$\leq 10^7$~cm is derived. This excludes the presence of an accretion disk 
in the case of a white dwarf. A disk could form outside a neutron star 
magnetosphere only for magnetic fields of 
$B \leq 3\times 10^8 (\dot M_x/10^{15}{\rm g~s^{-1}})^{1/2}$~G.

The galactic HI column in the direction of  HD49798 is $\sim 6\times
10^{20}$~cm$^{-2}$ (Stark \etal 1992). Uncertainties  
are introduced by the $\sim 2^{\rm o}$  radio beam-averaging. 
Moreover a fraction
of the column might be beyond HD49798,  despite its  height above the galactic
plane ($\sim 210$~pc). We conclude that the estimate above is not in
contradiction with the range of $\sim 1-3\times10^{20}$~cm$^{-2}$ derived
from the PSPC spectrum. 

At the descending node (close to the phase of the Rosat observation), 
the expected column density from a spherical wind is  
$N_{H,w}\simeq 5\times 10^{20} (\dot M_*/10^{-8}{\rm M_{\odot}~yr^{-1}}) 
(v_w/10^3 {\rm km~s^{-1}})^{-1}(a/5\times10^{11}{\rm cm})^{-1}$~cm$^{-2}$,
with $a$ the orbital separation. 
Abundances in the sdO star atmosphere 
are published only for He ($\times 2.8$ solar), C ($\times0.05$),  
N ($\times 25$) and Si ($\times 1.4$) (cf. Bauer \& Husfeld 1995). 
The absence of substantial O, Ne and Mg lines 
indicates that these elements are underabundant (probably 
$\times 0.1-0.3$, similar to other sdO stars).
Therefore in the Rosat PSPC energy range, He and N should 
dominate the photoelectric absorption by the wind (H is virtually 
100\% ionized); indeed  
given the UV flux from the sdO star the most populated stages are probably 
HeII and NIV-NVI. In the following we discuss the X-ray spectrum of 
HD49798 in the relation to the expected absorption properties of the wind. 

In the white dwarf interpretation, the required mass inflow rate 
implies that the wind optical depth at the HeII and NIV--VI edges is
$\geq 2000$ and $\geq 4$, respectively. 
Therefore for the upper limit on the HeII optical depth not to be violated
(see Sect.~2), the fraction of HeIII must be very large. 
The white dwarf accretion luminosity is insufficient to photoionize HeII. 
The required level of He ionization could be due to an additional 
wind heating mechanism (temperatures $\geq 10^5$~K), such as, possibly,  
the friction between a low and a high speed wind component 
(see Springmann \& Pauldrach 1992). 
In any case, a conspicuous N K-edge should be present around energies of
$0.46-0.55$~keV (ionization stages above He-like N would not be populated).
This, in turn, favors the spectral decomposition involving a 
power law and a N K-edge (see Sect.~2). 

If the accreting object is a neutron star, the observed 0.1-2~keV luminosity  
can be produced for smaller
mass inflow rates which do not require appreciable wind absorption from HeII 
and NIV--VI. Accretion luminosities of up to
$\sim 10^{35}$~erg~s$^{-1}$, as suggested by the extrapolation of the 
steep power law--like PSPC spectrum to the EUV, would still be compatible with 
the wind parameters. In this case the EUV luminosity could photoionize  
He to the required level, while a NIV--VI K--edge optical depth 
of a few could still be present.  
For bolometric luminosities in the $10^{32}-10^{35}$~erg~s$^{-1}$ range, 
the blackbody plus power law model implies $R_{bb} \sim 20-2000$~km 
(see Fig.\,2d) a difficult range to reconcile with a neutron star.

\subsection{White Dwarf accretor}

If the degenerate star in HD49798 is a white dwarf, it would be the first
example of a white dwarf accreting from an early type (though very
unusual !)  companion. The 13 s X--ray pulsations
might originate from magnetic polar cap accretion, as in the
case of intermediate polars (IPs). The 13~s rotation period would then
be the shortest known.  
By requiring that the magnetospheric radius is smaller than the 
corotation radius ($r_m < r_c$), such that the ``centrifugal barrier" 
is open and accretion can take place, 
we derive $B \leq 5 \times 10^3 (\dot M_x/10^{15}{\rm g~s}^{-1})^{1/2}$~G. 
This value is substantially lower  
than the  range inferred for IPs ($\sim 10^5-10^7$~G).
While most IPs display hard X--ray spectra, a small subgroup with spin periods
of 5--15~min is characterized by an additional very soft spectral component,
which probably originates from the reprocessing at the white dwarf surface of
the primary hard X--ray radiation emitted at the end of the accretion column(s)
(Duck \etal 1994; Haberl \& Motch 1995).  
The very soft X-ray spectrum of HD49798, however, must have a different
origin. Indeed in this case, $r_c \sim 8 \times 10^8$~cm sets an upper 
limit of $\sim 3$ white dwarf radii, $r_{wd}$, to $r_m$.
Accretion is therefore expected to take place over a large fraction of the
white dwarf surface, $f \sim r_{wd}/r_m \sim 1/3$
(see e.g. Frank, King \& Raine 1985). This is far larger than the $R_{bb}$ 
derived from blackbody fits for luminosities of $\sim 10^{32}$erg~s$^{-1}$
(see Sect.~2 and Fig.\,2d).

The nova-like variable AE~Aqr presents a composite soft X--ray  spectrum with
a luminosity of $\sim 10^{31}$~erg~s$^{-1}$ and  X--ray pulsations at the 33~s
white dwarf spin period,  currently the fastest known (Eracleous \etal 1991). 
The power implied by the measured spin-down rate far 
exceeds the UV and X--ray luminosity (de Jager \etal 1994).  This,
together with the absence of an accretion disk,  indicates that AE~Aqr hosts a
magnetic white dwarf in the ``propeller" regime, which expels most of the
inflowing matter, causing  large spin-down
torques (Wynn, King \& Horne 1996). It is unlikely that HD49798 hosts 
a similar propeller, as this would require a wind mass capture rate 
$\gg 10^{15}$~g~s$^{-1}$.

The limit on the coherence $Q$ of the $\sim$ 13~s signal in the Rosat light 
curve of HD49798 ($Q \geq  10^6$) suggests an analogy with 
the $\sim 10-30$~s quasi--coherent oscillations
($Q\sim 10^4-10^6$), that are seen at soft X--ray and EUV
energies during the outbursts of dwarf novae, such as SS~Cyg and U~Gem
(C\'ordova \etal 1980; Jones \& Watson 1992; Mauche 1996). 
Note that the outburst X-ray spectra of these system are 
very soft ($kT_{bb} $ $\sim 25-30$~eV). 
The origin of these oscillations is still debated, but they are
known to be associated to disk-accreting cataclysmic variables 
accreting at rates of $\sim 10^{15}$~g~s$^{-1}$. 
White dwarf disk accretion, however, is ruled out for HD49798.

\subsection{Neutron Star accretor} 

If HD49798 hosts an accreting magnetic neutron star, then the $\sim$ 13~s 
pulsations  
arise from its rotation. The higher energy conversion efficiency makes it 
easier to reconcile the mass capture rate inferred from the sdO star 
wind properties not only with the measured luminosity in the Rosat PSPC band,
but also with plausible extrapolations to the EUV. We note that in the 
single blackbody (plus N edge) interpretation of the PSPC spectrum, the 
inferred black body radius is consistent with a  neutron star. 
The condition that the centrifugal barrier is open requires  
$B \leq 6 \times 10^{11} (\dot M_x/10^{15}{\rm g~s}^{-1})^{1/2}$~G. 
An accretion disk cannot form unless $B$ is three orders of magnitude lower. 
HD49798 should therefore alternate spin--up and spin--down episodes, similar 
to those observed in wind-fed X--ray pulsars (cf. White \etal 1995).
The spectra of most X-ray pulsars are hard 
($\Gamma\sim 0-2$) and extend to energies of several 
tens of keV. However a group of 4-5 accreting pulsars with 5-9~s periods 
and very soft X-ray spectra ($\Gamma\sim 2-5$) has been recently identified
(Mereghetti \& Stella 1995). This group is also 
characterized by X-ray luminosities of $\sim 
10^{35}-10^{36}$~erg~s$^{-1}$ and
secular spin-down, indicating that the neutron 
stars are close to their equilibrium period and have 
a relatively low $B$ of $\sim 10^{11.5}-10^{12}$~G.
Clearly there are similarities with the case of HD49798. 
On the other hand these X-ray pulsars 
either have a very low mass companion, 
or are isolated and accrete from a residual disk
(van Paradijs \etal 1995). 

In this interpretation, HD49798 would be the first
X--ray binary with an early type H-depleted mass donor. Such system would 
likely be the remnant of a high mass X--ray binary after common
envelope and  spiral in, an evolutionary phase that has not been seen
before.  If the sdO star is too light to undergo 
gravitational collapse, 
the system will evolve in $\sim 10^6$~yr into a non-interacting binary 
consisting of a massive white dwarf and a neutron star. Even if mass 
transfer rate were maintained throughout this transition at its current 
maximum value, 
the neutron star could be spun-up only if $B \leq 3\times10^8$~G, 
achieving a minimum period of  $\sim 7\ (B/10^8~{\rm G})^{6/7}$~ms 
(i.e. the ``spin--up line" for 
$\dot M_x \sim 10^{15}$~g~s$^{-1}$).  The currently known millisecond
pulsars with massive white dwarf companions have  
substantially shorter spin periods (Bailes \etal 1994; Camilo \etal 1996):  
therefore a system like HD49798 could not be their progenitor.
 
\section{Conclusions}

Our results prove that HD49798 hosts  
either a white dwarf or, more likely, a magnetic neutron star.
The value and sign of any changes in the $\sim 13$~s X-ray period  
should clarify the nature of the accreting degenerate star. 
The dwarf nova oscillation scenario requires quasi--coherent oscillations. 
If instead the $\sim 13$~s pulsations arise directly from the rotation 
of the degenerate star, they would be coherent enough to make HD49798 
a ``double spectroscopic" binary; in this case the system would hold a great 
potential for accurate mass measurements of a previously unobserved  
evolutionary stage.

\acknowledgments

GLI and LS acknowledge useful discussions with F. D'Antona, 
F. Fiore, R. Sancisi and A. Tornamb\`e. This work was partially 
supported through ASI grants.

\newpage

\section*{Figure Captions} 

{\bf Figure 1:} Power spectrum of the 0.1--2.0 keV ROSAT PSPC light curve of 
HD49798. The  highest peak, centered around 0.076 Hz, corresponds to the
13.2~s pulsations.  The folded lightcurve is shown in the inner panel.

{\bf Figure 2:} Unfolded spectra and best fit models from the ROSAT PSPC 
observation of HD49798. Panel $a$ refers to the power law 
model (residuals are also shown), panel $b$ to the power law and NIV K-edge 
model and panel $c$ to the blackbody plus power law model.
Panel d shows the the 68.3\% , 95.4\% and 99.7\% 
confidence level contour plot in the $N_H$ -- $kT_{bb}$ 
plane for the model in panel c. 
Lines of constant black body radius $R$ (in km, dashed lines) 
and bolometric luminosity $L_{bb}$ (in erg~s$^{-1}$, solid lines) 
are also shown.

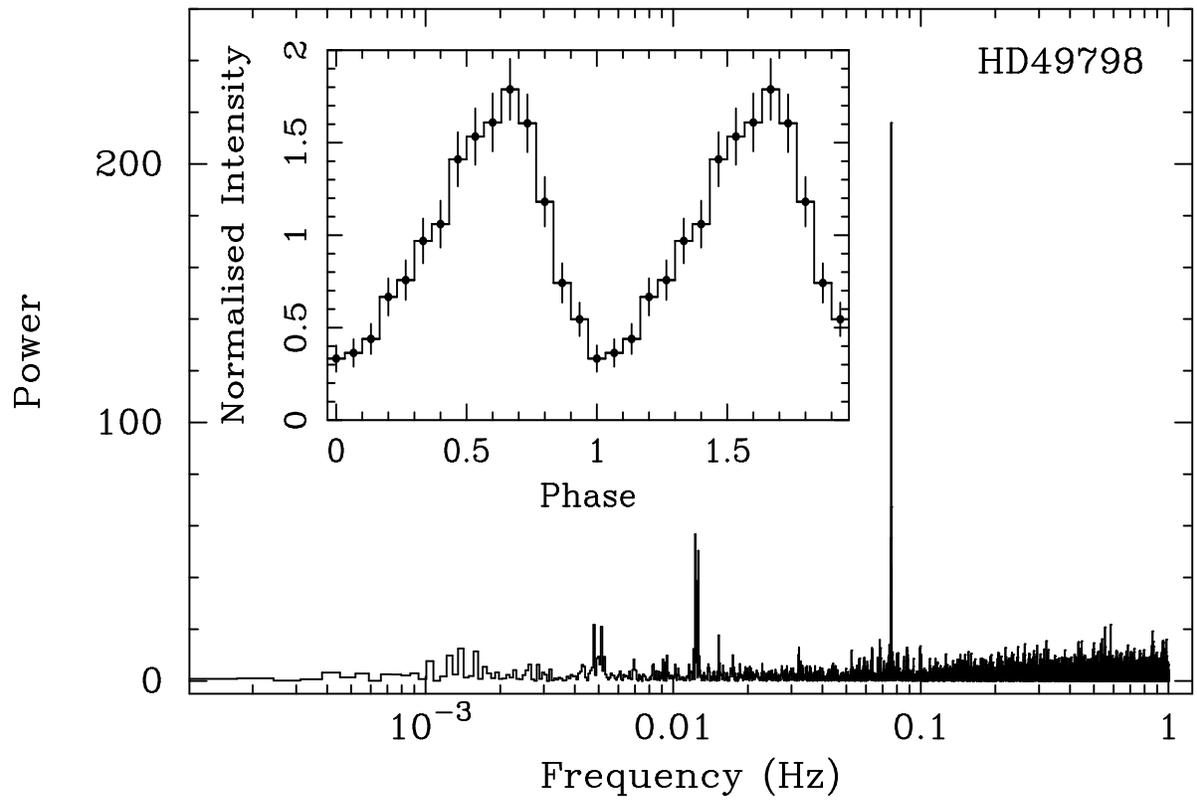
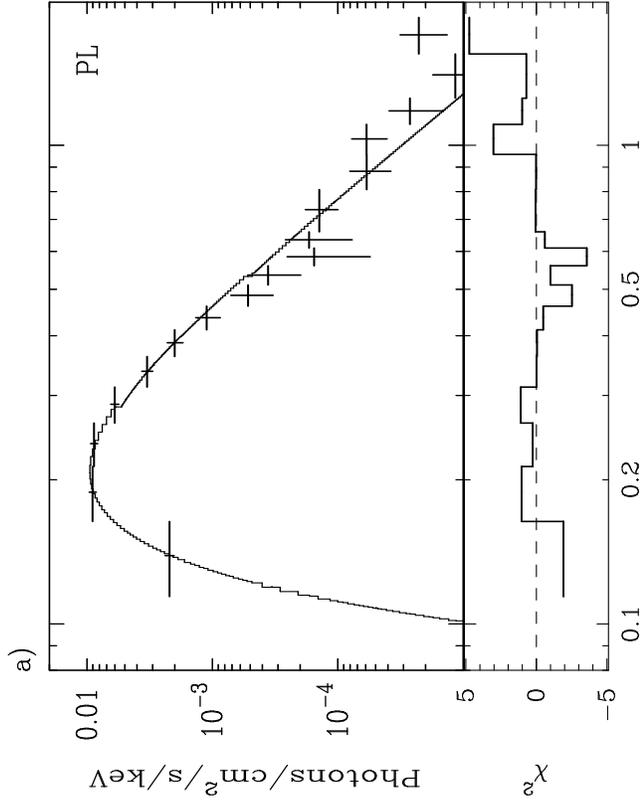
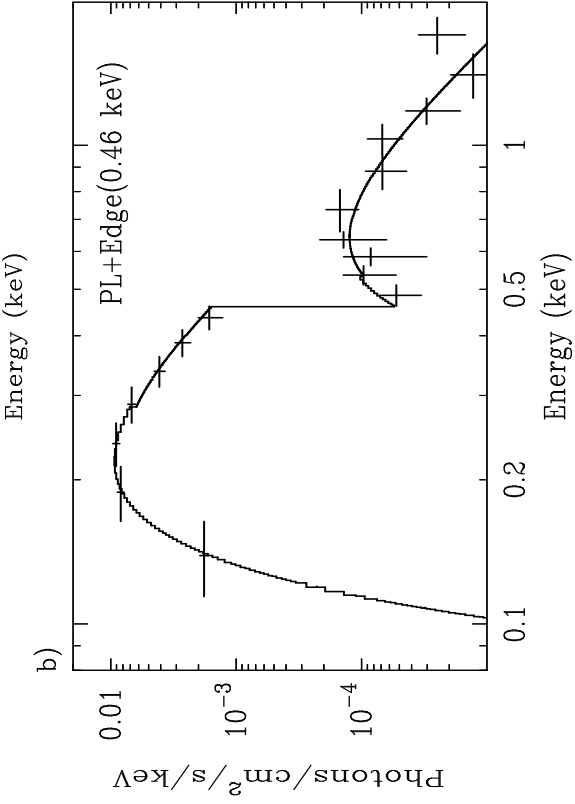
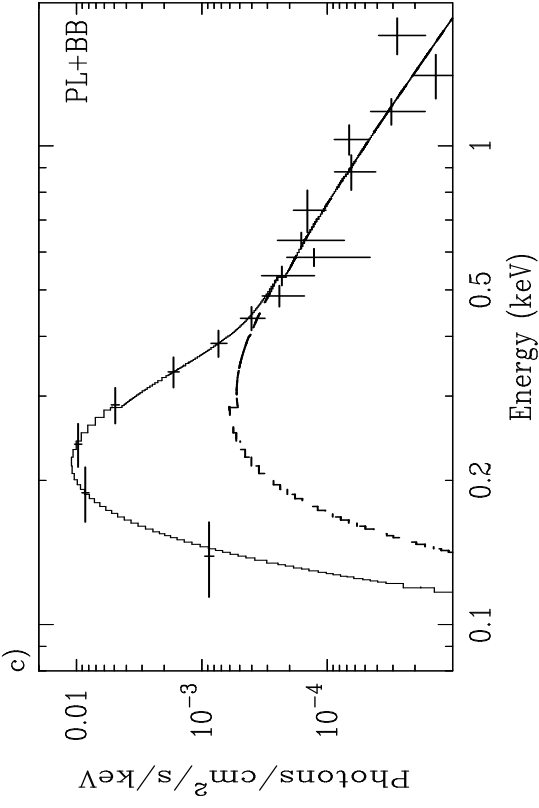
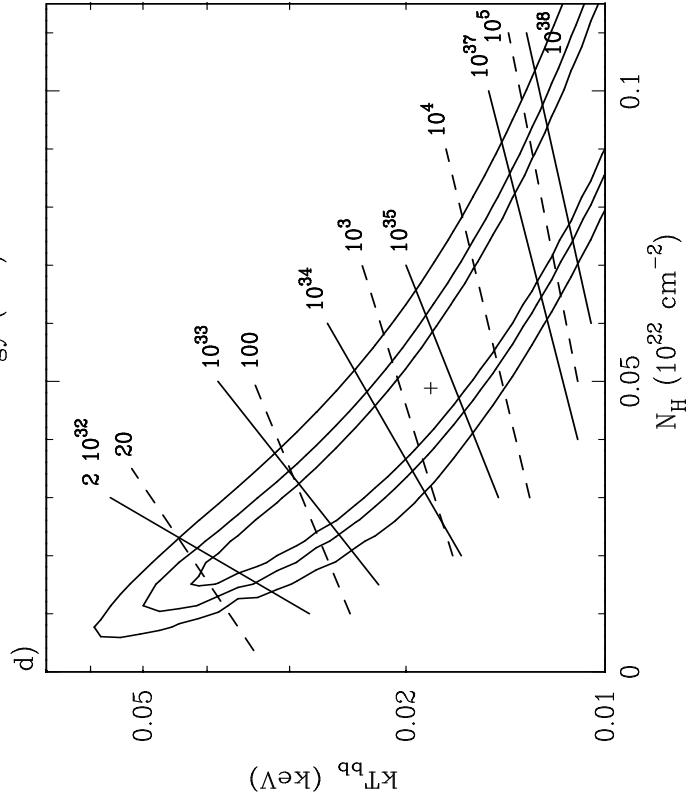
\end{document}